\documentclass[showpacs, showkeys,bf, preprint,english,11pt]{revtex4}
\usepackage{graphicx}
\usepackage{mathrsfs}
\usepackage[hypertex]{hyperref}
\usepackage{amssymb,amsmath}
\usepackage{bm}

\begin{document}

\title{$\Lambda$CDM and Power-Law Expansion in Lyra's Geometry}
\author{\textsc{Hoavo  Hova}\footnote{hovhoavhill@yahoo.fr}}

\affiliation{D\'{e}partement de physique, Facult\'{e} des Sciences et Techniques,\\ Universit\'{e} de Kara, Togo}

\begin{abstract}
We establish in a cosmological model based on Lyra's geometry a relationship between a displacement vector field $ \phi_{\mu} $, the Hubble parameter and the matter energy density $ \rho_{m} $ (with a constant equation of state $0 \le \omega_{m} < 1$) via an arbitrary function $\alpha(t)$. For a pressureless matter ($ \omega_{m}=0 $) the effective equation of state $ \omega_{eff} $ is completely determined by $ \alpha(t) $. We subsequently investigate and find exact solutions in models yielding the $ \Lambda $CDM and a power-law expansion.
\end{abstract}


\maketitle

\section{Introduction}
Since the discovery of an accelerated expansion of the universe from astrophysical observations \cite{Obs} and the impossibility of explaining this acceleration with the Einstein gravitational field equations of general relativity, $ R_{\mu\nu} - \frac{1}{2} g_{\mu\nu} R = T^{m}_{\mu\nu}$, cosmologists think that, beside the standard matter there must be an unknown fluid with a \emph{negative pressure} $ p_{\textsc{de}} < -\frac{1}{3} \rho_{\textsc{de}} $, that drives the universe into acceleration. Such a component, dubbed ``dark energy'' was completely negligible in most of the past and will entirely dominate in the future. The origin and the exact nature of dark energy are still unknown. Subsequently, several interesting models have considered beside the standard matter an \emph{alien dynamic} dark energy component  \cite{Ratra, Onemli, Caldwell, Armendariz, Padmanabhan, Li, Cai, Wei, Kamenshchik, Zhang} (for further reading see \cite{Copeland, Bamba} and references therein).

Attempting to solve geometrically the dark energy mechanism, a different approach considers the modification of the Einstein gravitational field equations with models involving only the standard matter.  All the additional terms in the \emph{modified Einstein field equations} beside the usual terms of general relativity constitute the so-called dark energy component. Such a modification can arise, either by extending the Einstein-Hilbert action to a more fundamental theory ($ f(R, R_{\mu\nu}, R_{\mu \nu \alpha \beta}) $ theories of gravity constructed in \emph{pseudo-Riemannian geometry}) or by modifying the Riemannian geometry. In the former case the simplest $ f(R) $ is that contains the \emph{cosmological constant} $\Lambda$ in the form $ f(R)= R-2 \Lambda $. The cosmological constant thus introduces a constant contribution to the total energy-momentum  tensor, which then has an effective equation of state $\omega_{eff}<-\frac{1}{3}$, leading hence to an accelerated expansion. In this framework, the $\Lambda$ dark energy is well-defined and reduces to a single number equivalent to a fluid with equation of state $\omega_{\Lambda} = p_{ \Lambda}/ \rho_{\Lambda} = -1$. Although the $\Lambda$ dark energy is indeed consistent with all our current cosmological data and does not introduce new degrees of freedom, it encounters, however, the fine-tuning and cosmic coincidence problems by the fact that it is a non-dynamical quantum vacuum energy and its observed value is much less than the theoretical value computed from quantum field theory, $ \rho_{obs} \approx 10^{-123} \rho_{vac}^{th}$. But several choices for $f(R, R_{\mu \nu},R_{\mu \nu \alpha \beta})$ have led to dynamic dark energy roughly compatible with observational data (for recent Review see \cite{Antonio, Nojiri} and references therein). Unfortunately, apart from $f(R)= R-2 \Lambda $ the modifications of the gravitational field equations from $f(R,R_{ \mu \nu},R_{\mu \nu \alpha \beta})$ theories include third or higher-order derivatives of metric components. Furthermore, higher-order terms in curvature invariants (such as $ R^{2} $, $ R_{ \mu \nu}R^{ \mu \nu}$, the \emph{Kretschmann scalar} $R_{ \mu \nu \alpha \beta}R^{ \mu \nu \alpha \beta}$ \cite{Kretschmann}, $ R \square R $ or $ R \square^{k} R $) or non-minimally coupled terms between scalar fields and geometry (such as $\varphi^{2}R$) have to be added in an \emph{ad hoc way} to the effective Lagrangian of gravitational field when quantum corrections are considered.

Another kind of modification of the Einstein gravitational field equations, mentioned above,  is that constructed within Lyra's geometry  \cite{Lyra}, particularly in the so-called \emph{normal gauge}. Lyra's geometry \cite{Lyra, Sen1, Sen2, Matyjasek, Matyjasek1}, like  Weyl's geometry \cite{Weyl}, is a modified Riemannian geometry proposed in order to unify gravitation and electromagnetism into a single spacetime geometry. However, in the framework of general relativity, only the connection in Lyra's and Riemannian geometries (contrary to Weyl's geometry) preserves the metric and the integrability of length transfers. Hence, Lyra's geometry becomes the most acceptable modified Riemannian geometry to describe the dynamic of the universe. Indeed, theories investigated in Lyra's geometry predict the same effects within observations limits, as far as the classical Solar System, as well as tests based on the linearized form of the field equations and are free of the Big-Bang singularity. They also solve the entropy and horizon problems, which beset the standard models based on Riemannian geometry \cite{Sen1, Halford1}. The modifications based on Lyra manifold are generally simpler than those built from  $f(R, R_{\mu\nu}, R_{\mu \nu \alpha \beta})$ theories, since they involve only first and second derivatives of metric components, and are basically due to the presence of an intrinsic geometrical \textit{vector field} $\phi_{\mu}$ (arising from the concept of a gauge) in the geometrically structureless manifold, without adding higher-order terms. On the other hand, the cosmological term (\emph{variable} or \emph{constant}), introduced in an ad hoc manner into the gravitational field equations has in Lyra's geometry an intrinsic geometrical origin arising from the displacement field \cite{Halford, Beesham, hova}. In addition, the scalar-tensor theories of gravitation constructed within Lyra's geometry involve \emph{a scalar field and a tensor field that are all intrinsic to the geometry} \cite{Sen1, Sen2, Halford1, Halford,  Beesham,  Sen3, Sen4, Scheibe, Soleng, Soleng1, Hudgin, Beesham1}.
 
In this work we are to establish through an arbitrary function $\alpha(t)$ a connection between the Hubble parameter, the displacement field and the matter energy density, and express the effective equation of state in terms of $\alpha(t)$ and $\omega_{m}$. Considering a pressureless matter we also study a possible accelerated expansion of the universe, for a time-dependent and a constant functions $\alpha(t)$.

The paper is organized as follows. In Section II, we describe a model, where a relationship between the Hubble parameter, the displacement field and the matter energy density is established, and we also investigate the $\Lambda$CDM model and Power-Law Expansion. Conclusions are given in Section III.

\section{The Model}
In \emph{Planckian units} $ c = \hbar = \kappa^{2} = 1 $, the Einstein gravitational field equations in normal-gauge Lyra manifold as obtained in \cite{Sen1} are:
\begin{equation}
\label{h}
G_{ \mu \nu} + \tau_{\mu \nu} = T_{\mu \nu},
\end{equation}
where $ G_{\mu \nu} = R_{\mu \nu} - \frac{1}{2} g_{\mu \nu} R $ is the Einstein tensor ($ R_{\mu \nu} $ and $ R $ being Ricci tensor and Ricci scalar, respectively),
\begin{equation}
\tau_{\mu \nu} = \frac{3}{2} \left(\phi_{\mu} \phi_{\nu} - \dfrac{1}{2} g_{\mu \nu} \phi_{\lambda} \phi^{\lambda} \right)
\end{equation}
is an  intrinsic geometrical tensor associated with the displacement vector field $\phi_{\mu}$, whereas $T_{\mu \nu}$, defined by
\begin{equation}
T_{\mu \nu} = \left( p_{m} + \rho_{m} \right) u_{\mu} u_{\nu} +p_{m} g_{\mu \nu}
\end{equation}
 is the energy-momentum tensor of a perfect fluid with energy $\rho_{m}$ and pressure $p_{m}=\omega_{m}\rho_{m}$ ($0\le \omega_{m}< 1$). The vector $u_{\mu}=(1,0,0,0)$, with  $u^{\mu}u_{\mu}=-1$, is the 4-velocity of the comoving observer. When the tensor $\tau_{\mu\nu}$ vanishes, we recover the gravitational field equations of general relativity; so the modification of the Einstein's equations is solely due to the intrinsic geometrical vector field $\phi_{\mu}$.
 
 In the flat \textit{Friedmann-Lema\^{i}tre-Robertson-Walker} background $ds^{2} = -dt^2 + a^{2} \big( dr^{2} + r^{2} d\theta^{2} + r^{2} \sin^{2} \theta d\varphi^{2} \big)$, where $a = a(t)$ is the scale factor of an expanding universe, the nonvanishing components of the Ricci tensor, and the Ricci scalar are given by
 \begin{equation}
 R_{00} = -3\big(\dot{H}+H^{2}\big), \quad R_{ij} = \big(\dot{H}+3H^{2}\big)g_{ij}, \quad R=6\big(\dot{H}+2H^{2}\big),
 \end{equation}
where an overdot denotes differentiation with respect to the time coordinate $t$ and $H=\dot{a}/a$ is the Hubble parameter. Throughout this paper we are to consider a time-like displacement vector 
\begin{equation}
\phi_{\mu}=\left(\phi(t),0,0,0 \right).
\end{equation}
Thus, the gravitational field equations (\ref{h}) reduce to the Friedmann equations
\begin{eqnarray}
\label{2}
3H^{2}+\frac{3}{4}\phi^{2}(t)=\rho_{m},\\
\label{3}
-\left( 2\dot{H}+3H^{2}\right) +\frac{3}{4}\phi^{2}(t) =\omega_{m}\rho_{m},
\end{eqnarray}
while applying the Bianchi identities we find the total energy conservation equation
\begin{equation}
\label{4}
\dot{\rho}_{m}+3H(1+\omega_{m})\rho_{m} = \frac{3}{2}\left(\phi\dot{\phi}+3H\phi^{2} \right).   
\end{equation}
The matter energy interacts with the displacement field as the universe evolves. In \cite{hova} an interaction was built between $\phi$ and the matter energy $\rho_{m}$ by encoding  Eq. (\ref{4}) into two conservation equations. Here, we are to consider the first Friedmann equation (\ref{2}) to find the connection between $\phi$, $H$ and $\rho_{m}$. Assuming therefore the geometrical quantities $H$ and $\phi$ are both real, we can recast Eq. (\ref{2}) in the form
\begin{equation}
3\big(H+\dfrac{i}{2}\phi\big)\big(H-\dfrac{i}{2}\phi\big) = \rho_{m}
\end{equation}
or compactly
\begin{equation}
\label{1}
Z\bar{Z} = |Z|^{2} = 1,
\end{equation}
where the dimensionless quantities $Z$ and  $\bar{Z}$ are two c-numbers, functions of time $t$, defined by
\begin{equation}
Z = \sqrt{\dfrac{3}{\rho_{m}}}\big(H+\dfrac{i}{2}\phi\big) \qquad \mbox{and} \qquad \bar{Z} = \sqrt{\dfrac{3}{\rho_{m}}}\big(H-\dfrac{i}{2}\phi\big) \qquad \mbox{with} \qquad i=\sqrt{-1}.
\end{equation}
In terms of $Z$ and its conjugate $\bar{Z}$ the Hubble parameter and the displacement field read
\begin{equation}
H=\sqrt{\dfrac{\rho_{m}}{3}}~\dfrac{Z+\bar{Z}}{2}
\end{equation}
and
\begin{equation}
\phi=2\sqrt{\dfrac{\rho_{m}}{3}}~\dfrac{Z-\bar{Z}}{2i},
\end{equation}
respectively. As for every c-number the general form of $Z(t)$ can be parametrized as 
\begin{equation}
\label{key0}
Z(t)=\beta (t)\mbox{e}^{i\alpha(t)},
\end{equation} 
where $\alpha (t)$ and $\beta (t)$ are two arbitrary real functions of time. Hence, $H$ and $\phi$  become
\begin{equation}
\label{key}
H(t)=\dfrac{1}{\sqrt{3}}\beta (t) \sqrt{\rho_{m}(t)} \cos[\alpha(t)]
\end{equation}
while
\begin{equation}
\label{key1}
\phi(t)=\dfrac{2}{\sqrt{3}}\beta (t) \sqrt{\rho_{m}(t)} \sin[\alpha(t)].
\end{equation}
To reduce furthermore Eqs. (\ref{key}) and (\ref{key1}), we substitute them into (\ref{2}) or use straightforwardly the identity (\ref{1}) ($Z\bar{Z} = 1$), and we find $\beta^{2} (t) = 1$. 
That leaves therefore $\alpha(t)$ as the only unknown function characterizing the c-number $Z$. Observationally, the Hubble parameter $H$ and the energy density $\rho_{m}$ are positive; so taking $\beta(t)=+1$ (from now on) the sign of $H$ is determined by that of $\cos[\alpha(t)]$, and we must have $\cos[\alpha(t)]>0$ whatever the sign of $\sin[\alpha(t)]$ ($\phi$ can be positive or negative). Note also that, depending on $\sin[\alpha(t)]$, the displacement field $\phi$  could vanish yielding a decelerating expanding universe. But our goal is to explain the recent cosmic acceleration, thus we require $\sin[\alpha(t)] \ne 0$. 

Moreover, combination of Eqs. (\ref{key}) and (\ref{key1}) relates $H$ to $\phi$ through the equation
\begin{equation}
\label{key2}
\phi(t) = 2 \tan[\alpha(t)] H(t).
\end{equation}
Thus, $\alpha (t)$ appears as playing the role of interaction between $H$, $\phi$ and $\rho$, and will be called ``\emph{interaction-function}'' from now on.
 
On the other hand, in view of (\ref{key2}), equation (\ref{3}) transforms according to 
\begin{equation}
\label{key3}
2\dot{H}(t)+3\big(1-\tan^{2}[\alpha(t)]\big)H^{2}(t)= - \omega_{m} \rho_{m}(t)
\end{equation} 
while the energy conservation equation (\ref{4}) becomes
\begin{equation}
\label{key4}
\dot{\rho}_{m}(t) +3 H \big[1+\omega_{m} - (1-\omega_{m})\tan^{2}[\alpha(t)]\big] \rho_{m} (t) = 2\dot{\alpha}(t) \tan[\alpha(t)]\rho_{m} (t). 
\end{equation}
Using the e-folding number $x = \ln a$, Eq. (\ref{key4}) may be written as a conservation equation of the matter energy $\rho_{m}$ with an ``\emph{effective equation of state}'' $\tilde{\omega}_{m}(x)$ in the form
\begin{equation}
\label{key5}
\dfrac{d\rho_{m}(x)}{dx} +3 \big[1 + \tilde{\omega}_{m}(x) \big]\rho_{m} (x) = 0,
\end{equation}
where 
\begin{equation}
\label{key6}
\tilde{\omega}_{m}(x) \equiv \omega_{m} - (1-\omega_{m})\tan^{2}[\alpha(x)]-\dfrac{2}{3}\tan[\alpha(x)]\dfrac{d\alpha(x)}{dx}.
\end{equation}
It is important to emphasize that, even if  $\omega_{m}=0$, the matter effective equation of state   $\tilde{\omega}_{m}(x)$ is nonzero, unless the interaction-function $\alpha (x)$ varies as
\begin{equation}
\label{0}
\alpha(x) =\left\lbrace \begin{array}{rlll}
n\pi, \quad \omega_{m}=0 \quad \mbox{with} \quad  n~\epsilon~ \mathbb{Z},~~~~~~~~~~~~~~~~~~~~~~~~~~~~~~~~~~~~~~~~\\
\arcsin\big(k~\mbox{e}^{-\frac{3}{2}x}\big) , \quad \omega_{m}=0 \quad \mbox{with} \quad x \ge \dfrac{2}{3}\ln\big(|k|\big),~~~~~~~~~~~~~~~~~~~~~~~~~~~~\\
\arcsin\big(\sqrt{\omega_{m} + \tilde{k}~\mbox{e}^{-3x}}\big), \quad  \omega_{m}\ne 0 \quad \mbox{with} \quad \left\lbrace \begin{array}{rlll} x \ge \dfrac{1}{3}\ln\bigg(\dfrac{\tilde{|k|}}{\omega_{m}}\bigg) \quad \mbox{if}\quad \tilde{k}<0, \\
x \ge \dfrac{1}{3}\ln\bigg(\dfrac{\tilde{k}}{1-\omega_{m}}\bigg) \quad \mbox{if}\quad \tilde{k}>0.
 \end{array}\right.
\end{array}\right.
\end{equation}
 $k$ and $\tilde{k}$  are constants of integration.
 
The matter effective equation of state $\tilde{\omega}_{m}(x)$ is not the \emph{effective equation of state} of the system, $\omega_{eff}$, which by contrast is defined by
\begin{equation}
\omega_{eff}(t)\equiv -1-\dfrac{2}{3}\dfrac{\dot{H}}{H^{2}}. 
\end{equation}
By using (\ref{key}) and (\ref{key3}) knowing that $\beta(t)=+1$, $\omega_{eff}(t)$ reduces to
\begin{eqnarray} 
\label{key7}
\omega_{eff}(t)= \omega_{m} - (1-\omega_{m})\tan^{2}[\alpha(t)].
\end{eqnarray}
or equivalently,
\begin{equation}
\omega_{eff}(x)=\tilde{\omega}_{m}(x) + \frac{2}{3}\tan[\alpha(x)]\frac{d\alpha(x)}{dx}.
\end{equation}
Which shows for $\tan[\alpha(t)] \ne 0$ that the difference between $\tilde{\omega}_{m}$  and $\omega_{eff}$ comes from the variation of the interaction-function. For a pressureless matter ($\omega_{m} = 0$), $\tilde{\omega}_{m}$ and $\omega_{eff}$ take the simple forms
\begin{equation}
\label{key8}
\tilde{\omega}_{m}(x) =  - \tan^{2}[\alpha(x)]-\dfrac{2}{3}\tan[\alpha(x)]\dfrac{d\alpha(x)}{dx}
\end{equation}
and
\begin{equation} 
\label{key9}
\omega_{eff}(t)= - \tan^{2}[\alpha(t)],
\end{equation}
respectively. So, when $\omega_{m} = 0$ both $\tilde{\omega}_{m}$  and $\omega_{eff}$ may be completely determined by the interaction-function $\alpha(t)$ only. Assuming a positive $H$ (then $\cos[\alpha(t)]>0$) and any $\phi\ne 0$ acceleration occurs (for $\omega_{m}=0$) when $\omega_{eff}(t)= - \tan^{2}[\alpha(t)]<-\frac{1}{3}$, i.e. for $\alpha(t)$ varying in the ranges
\begin{equation}
-\dfrac{\pi}{2}<\alpha(t)-2n\pi < -\dfrac{\pi}{6} \quad \mbox{or} \quad \dfrac{\pi}{6}<\alpha(t)-2n\pi < \dfrac{\pi}{2} \quad \mbox{with} \quad
 n~\epsilon~ \mathbb{Z}.
\end{equation}

In the following we are to consider a time-evolving and a constant interaction-functions and study then their consequences on the dynamics of the universe.

\subsection{$\Lambda$CDM model}

For $\omega_{m} = 0$ and taking 
\begin{equation}
\label{key10}
\alpha(x)=\arctan\big[(1+\lambda \mbox{e}^{-3x})^{-1/2}\big]
\end{equation}
with $\lambda$ a positive constant, leads as shown in \cite{hova} to the $\Lambda$CDM model and a constant displacement field $\phi_{0}$, that plays the same role as the \emph{cosmological constant}: $\Lambda=\frac{3}{4}\phi_{0}^{2}$. It was found  that \cite{hova}
\begin{equation}
\label{eq}
\rho_{m} =\rho_{\textsc{cdm}} +2\Lambda, \quad 3H^{2} = \rho_{\textsc{cdm}} + \Lambda \quad \mbox{with} \quad \rho_{\textsc{cdm}} = \rho^{0}_{\textsc{cdm}}a^{-3},
\end{equation}
where $\rho^{0}_{\textsc{cdm}}$ is the present CDM energy density (that is, at $t=t_{0}$ or $x=0$).  Notice that Eq. (\ref{key10}) can be computed straightforwardly by considering a constant displacement field $\phi(t)=\phi_{0}>0$ in (\ref{key2}) and using (\ref{key3}). One finds
\begin{equation}
\alpha(x)=\dfrac{1}{2}\arccos\bigg(\dfrac{1}{1+\zeta \mbox{e}^{3x}}\bigg)
\end{equation}
with $\zeta$ being a constant of integration. After some algebras we will arrive at Eq. (\ref{key10}) with the identity $\lambda=2/\zeta$.   

Now we are to find explicitly the time-dependence of $\alpha(t)$ and quantities in the $\Lambda$CDM model. Using then (\ref{key2}) and (\ref{key3}) together with the identity $\Lambda=\frac{3}{4}\phi_{0}^{2}$ gives
\begin{equation}
\label{keyy}
\alpha(t) = \arctan \bigg(1-\dfrac{2}{1+\tilde{\lambda}\mbox{e}^{\sqrt{3\Lambda}t}}\bigg),
\end{equation}
where $\tilde{\lambda}$ is a constant of integration. Substitution of (\ref{keyy}) into (\ref{key2}) and using the fact that $H \to +\infty$ at $t=0$ (yielding $\tilde{\lambda}=1$) enable one to obtain, 
\begin{equation}
H(t) = \sqrt{\Lambda/3}~\dfrac{\mbox{e}^{\sqrt{3\Lambda}t}+1}{\mbox{e}^{\sqrt{3\Lambda}t}-1}=\sqrt{\Lambda/3}\coth\big(\sqrt{3\Lambda/4}t\big)
\end{equation}
and
\begin{equation}
\label{eq1}
a(t)=a_{i}\big(1-\mbox{e}^{-\sqrt{3\Lambda}t}\big)^{2/3}~\mbox{e}^{\sqrt{\Lambda/3}t}= 2^{2/3}a_{i}\sinh^{2/3}\big(\sqrt{3\Lambda/4}t\big)
\end{equation}
with $a_{i}$ a constant of integration. Also, $\omega_{eff}$, $\tilde{\omega}_{m}$ and $\rho_{m}$ are evaluated as 
\begin{equation}
\omega_{eff}(t) = -\bigg(\dfrac{\mbox{e}^{\sqrt{3\Lambda}t}-1}{\mbox{e}^{\sqrt{3\Lambda}t}+1}\bigg)^{2}=-\tanh^{2}\big(\sqrt{3\Lambda/4}t\big),
\end{equation}
\begin{equation}
\tilde{\omega}_{m}(t)=-1 + \dfrac{2\mbox{e}^{\sqrt{3\Lambda}t}}{1+\mbox{e}^{2\sqrt{3\Lambda}t}}=-1+\mbox{sech}\big(\sqrt{3\Lambda}t\big)
\end{equation}
and
\begin{equation}
\label{e.q}
\rho_{m}(t)=2\Lambda~\dfrac{\mbox{e}^{2\sqrt{3\Lambda}t}+1}{\big(\mbox{e}^{\sqrt{3\Lambda}t}-1\big)^{2}}=\Lambda\cosh\big(\sqrt{3\Lambda}t\big)\mbox{csch}^{2}\big(\sqrt{3\Lambda/4}t\big). 
\end{equation}
Eq. (\ref{e.q}) shows that $\rho_{m}(t) \to 2\Lambda$ when $t$ becomes very large, and only the cosmological constant contributes to the matter energy $\rho_{m}(t)$, as seen from (\ref{eq}).

On the other hand, Eqs. (\ref{eq}) and (\ref{eq1}) allow one to compute the CDM energy $\rho_{\textsc{cdm}}$ and the constant of integration $a_{i}$:
\begin{equation}
\rho_{\textsc{cdm}}(t) = \Lambda \mbox{csch}^{2}\big(\sqrt{3\Lambda/4}t\big)
\end{equation}
and
\begin{equation}
a^{3}_{i}= \dfrac{\rho^{0}_{\textsc{cdm}}}{4 \Lambda} = \dfrac{\Omega^{0}_{\textsc{cdm}}}{4 \big(1-\Omega^{0}_{\textsc{cdm}}\big)},
\end{equation}
where $\Omega^{0}_{\textsc{cdm}}=\rho^{0}_{\textsc{cdm}}/3H^{2}_{0}$ is the present fractional density of CDM.

\subsection{Power-law expansion}

Here, we are going to study the case ``$\omega_{eff}(t)=\tilde{\omega}_{m}(t)$'' by considering a nonzero constant interaction-function, $\alpha(t)=\alpha_{0}$. This  leads for $\omega_{m} = 0$ to a \emph{power-law expansion} (derived from (\ref{key3})):
\begin{equation}
\label{key11}
H(t) = \dfrac{2}{3[1-\tan^{2}(\alpha_{0})]}~\dfrac{1}{t} \qquad \mbox{or} \qquad a(t)\propto t^{2/[3(1-\tan^{2}(\alpha_{0}))]}.
\end{equation}
Other cosmological quantities will be given by
\begin{equation}
\label{key12}
  \omega_{eff}(t)=\tilde{\omega}_{m}(t)= - \tan^{2}(\alpha_{0}), \qquad \phi(t) = \dfrac{4\tan(\alpha_{0})}{3[1-\tan^{2}(\alpha_{0})]}~\dfrac{1}{t}  
\end{equation}
\begin{equation}
\mbox{and} \qquad \rho_{m}(t) = \dfrac{4}{3}\dfrac{1+\tan^{2}(\alpha_{0})}{[1-\tan^{2}(\alpha_{0})]^{2}}~\dfrac{1}{t^{2}}. 
\end{equation}
Assuming $H(t)$ and $\phi(t)$ are all positive requires the conditions $\tan(\alpha_{0})>0$ and $\tan^{2}(\alpha_{0})<1$, what is translated by
\begin{equation}
\label{key13}
-\pi <\alpha_{0}-2n\pi<-\dfrac{3\pi}{4}   \qquad \mbox{or} \qquad  0 <\alpha_{0}-2n\pi<\dfrac{\pi}{4} \qquad \mbox{with} \qquad n~\epsilon~ \mathbb{Z}.
\end{equation}
In addition, the cosmic age problem is alleviated when together with the constraints (\ref{key13}) $\alpha_{0}$ also satisfies $2/[3(1-\tan^{2}(\alpha_{0}))]>1$, that is,
\begin{equation}
\label{key14}
-\dfrac{5\pi}{6} <\alpha_{0} -2n\pi<-\dfrac{3\pi}{4} \qquad \mbox{or} \qquad \dfrac{\pi}{6} <\alpha_{0} -2n\pi <\dfrac{\pi}{4} \qquad \mbox{with} \qquad n~\epsilon~ \mathbb{Z}.
\end{equation}
Hence, as the $\Lambda$CDM model, the $\alpha_{0}$-\emph{model} may satisfy the stellar bound on the age  estimation \cite{Jimenez, Richer, Hansen} and is free of the problem of the age of the  universe \cite{Spergel}. The conditions (\ref{key14}) are exactly those yielding an accelerated expansion of the universe.

\section{Conclusions}
In this work we constructed within Lyra's geometry a cosmological model involving a matter energy, with equation of state $0\le \omega_{m}<1$, interacting with the displacement field. After establishing a connection between the Hubble parameter and the displacement field through an interaction-function, we showed that the conservation equation of the total energy in the universe (matter + displacement field) could be written as that of the matter energy with an effective equation of state. Having considered, furthermore, a pressureless matter we studied an accelerated expansion  for a time-dependent and  a constant interaction-functions, leading to the $\Lambda$CDM model and a power-law expansion, respectively.


\providecommand{\href}[2]{#2}\begingroup\raggedright\endgroup

\end{document}